\documentclass{jstatphys}
\usepackage[reqno,sumlimits]{amsmath}
\usepackage{amsfonts}
\usepackage{amssymb}
\usepackage{bbm}
\def\rz{{\mathbbm{R}}}
\def\d{\mathrm d}
\def\e{\varepsilon}
\def\es{{\varepsilon^*}}
\begin{document}

\title{Consistency of Microcanonical and Canonical Finite-Size Scaling}

\author{M.~Kastner%
\footnotemark
\addtocounter{footnote}{-1}
and M.~Promberger%
\footnote{Institut f\"ur Theoretische Physik, Friedrich-Alexander-Universit\"at Erlangen-N\"urnberg, Staudtstra{\ss}e 7, 91058 Erlangen, Germany.}
}
\runningauthor{Kastner and Promberger}

\date{Version of \today}


\begin{abstract}
Typically, in order to obtain finite-size scaling laws for quantities in the microcanonical ensemble, an assumption is taken as a starting point. In this paper, consistency of such a Microcanonical Finite-Size Scaling Assumption with its commonly accepted canonical counterpart is shown, which puts Microcanonical Finite-Size Scaling on a firmer footing.
\end{abstract}

\section{Introduction}
Investigations of critical phenomena are focused on the properties of systems of infinite size. The application of computer simulational methods, however, supplies data for systems of finite size, and hence the need arises to have an extrapolation method which allows to extract information about the critical behaviour of the infinite system from finite system data. A prominent example of such an extrapolation method is Finite-Size Scaling as introduced by Fisher and Barber.\cite{FisherBarber} This method allows to determine critical exponents of the infinite system from the system-size dependence of certain canonical quantities of finite systems, and is therefore referred to as {\em Canonical Finite-Size Scaling}\/ ({\bf CFSS}) in the following.

Conventionally, the canonical approach is favoured for the investigation of phase transitions. Recent results, however, reveal advantages of the microcanonical ensemble, at least for the detection and classification of phase transitions\cite{DetClass} as well as for the localization of critical points\cite{Deserno} from finite system data. It is, {\em inter alia}, this fact which motivated the investigation of critical phenomena in the microcanonical ensemble and, to this purpose, the development of a {\em Microcanonical Finite-Size Scaling}\/ ({\bf MFSS}) theory.

The existing papers on MFSS\cite{Desai,KaProHue,BruWil,KastnerDiss} of various authors, although quite similar in their titles, are somewhat difficult to compare, which is in particular due to the fact that the naming ``microcanonical'' is used for different scenarios.%
\footnote{Bruce and Wilding,\cite{BruWil} for example, define their so-called ``microcanonical entropy density'' via a second order differential equation (!) from the reduced microcanonical partition function (\ref{Ored_def}).}
Typically,\cite{Desai,KaProHue} in analogy to CFSS, an assumption is taken as a starting point to derive MFSS laws which enable the determination of critical exponents of the infinite system from the system-size dependence of certain microcanonical quantities of finite systems. In this paper, it is shown that the MFSS Assumption as discussed in refs.\ \cite{KaProHue,KastnerDiss} is consistent with its canonical counterpart in the sense that validity of MFSS implies validity of CFSS.

\section{Notation}\label{notation}
We consider classical statistical systems of hypercubic geometry in $d$ spatial dimensions with volume $L^d$, where $L$ is the linear size of the system. Our interest is focused on systems which undergo a continuous equilibrium phase transition in the thermodynamic limit.

For notational simplicity, we restrict ourselves to Hamiltonians $\mathcal{H}$ of the form
\begin{equation}
\mathcal{H}:\;\Gamma\rightarrow\rz,\qquad x\mapsto L^d \left[\e(x)-h m(x)\right]
\end{equation}
where $\Gamma$ is the configuration space of the system and $h$ is an external magnetic field. $\e:\,\Gamma\rightarrow\rz$ and $m:\,\Gamma\rightarrow\rz$ map elements from configuration space onto their intensive interaction energy and magnetization value, respectively. We consider partition functions which depend on 2+1 variables,%
\footnote{The extension to more variables is straightforward. The reduction of the results to 1+1 variables is explicitely demonstrated in the Appendix.}
whereof one is the inverse linear system size $L^{-1}$. The microcanonical partition function or density of states
\begin{equation}
\Omega\,=\,\Omega(\es,m,L^{-1})
\end{equation}
is written as a function of the intensive magnetization $m$ and of the reduced interaction energy
\begin{equation}
\es\,:=\,\e-\e_c
\end{equation}
where $\e_c$ is the critical interaction energy. From the microcanonical partition function $\Omega$, the microcanonical entropy
\begin{equation} \label{sdef}
s^{mic}(\es,m,L^{-1})\,:=\,L^{-d}\ln\Omega(\es,m,L^{-1})
\end{equation}
is obtained.
The canonical partition function
\begin{equation} \label{Zdef}
Z(t,h,L^{-1})\,:=\,L^{2d}\int\d\es\d m \;\Omega(\es,m,L^{-1})\exp\left\{\frac{L^d (hm-\es-\e_c)}{T_c(1+t)}\right\}
\end{equation}
is a Laplace transform of $\Omega$ where the range of integration is $\rz^2$.
The canonical partition function is written as a function of the external magnetic field $h$ and of the reduced temperature
\begin{equation}
t\,:=\,\frac{T-T_c}{T_c}
\end{equation}
where $T$ is the temperature. For convenience, here and in the following we set Boltzmann's constant $k_B\equiv1$. From the canonical partition function, the canonical Gibbs free energy
\begin{equation} \label{gdef}
g^{can}(t,h,L^{-1})\,:=\,-T_c(1+t)L^{-d}\ln Z(t,h,L^{-1})
\end{equation}
is obtained.

In the following, we assume spin inversion symmetry such that the critical magnetization $m_c\equiv0$ and the critical external magnetic field $h_c\equiv0$. If this is not the case, $m$ and $h$ have to be substituted by appropriate reduced variables in the homogeneity relations of the next section.

\section{Finite-Size Scaling and Homogeneous Functions}\label{fss}
{\em Homogeneity relations}\/ can be taken as convenient starting points to obtain finite-size scaling laws.

Let us consider the standard case of CFSS first. From renormalization group arguments\cite{Suzuki,Brezin,Barber} it can be substantiated%
\footnote{Although recent results of Chen and Dohm\cite{ChenDohm} question the general validity of Br\'ezin's\cite{Brezin} arguments.}
that the canonical Gibbs free energy can be split into a regular and a singular part
\begin{equation}\label{g_split}
g^{can}(t,h,L^{-1})\,=\,g^{can}_{reg}(t,h,L^{-1})\,+\,g^{can}_{sing}(t,h,L^{-1})\;,
\end{equation}
where the second term in is subject to the homogeneity relation
\begin{equation}\label{g_hom}
\fbox{$\displaystyle \phantom{\int}g^{can}_{sing}(t,h,L^{-1})\,\simeq\,\lambda^{-1}g^{can}_{sing}(\lambda^{a_t}t,\lambda^{a_h}h,\lambda^{1/d}L^{-1})\;, \phantom{\int}$}
\end{equation}
valid asymptotically in the vicinity of the critical point $(t,h,L^{-1})=(0,0,0)$ for arbitrary positive values of the parameter $\lambda$. The exponents $a_t$ and $a_h$ determine the static universality class of the corresponding infinite system. The homogeneity relation (\ref{g_hom}) will be referred to as the {\bf CFSS Assumption} in the following, as it can serve as a starting point to derive scaling laws which describe the finite-size scaling behaviour of various canonical quantities (see e.g.\ ref.~\cite{Goldenfeld}).

In contrast to CFSS, its microcanonical counterpart is not a standard textbook formalism. To obtain a starting point for MFSS, it is argued\cite{Desai,KaProHue} that similar relations as in the canonical case should hold for microcanonical quantities. In analogy to Eqs.\ (\ref{g_split}) and (\ref{g_hom}), we choose the following formulation which will be referred to as the {\bf MFSS Assumption}: The microcanonical entropy can be split into  a regular and a singular part
\begin{equation}\label{s_split}
s^{mic}(\es,m,L^{-1})\,=\,s^{mic}_{reg}(\es,m,L^{-1})\,+\,s^{mic}_{sing}(\es,m,L^{-1})\;,
\end{equation}
where the second term is subject to the homogeneity relation
\begin{equation}\label{s_hom}
\fbox{$\displaystyle \phantom{\int}s^{mic}_{sing}(\es,m,L^{-1})\,\simeq\,\lambda^{-1}s^{mic}_{sing}(\lambda^{a_\e}\es,\lambda^{a_m}m,\lambda^{1/d}L^{-1})\;, \phantom{\int}$}
\end{equation}
valid asymptotically in the vicinity of the critical point $(\es,m,L^{-1})=(0,0,0)$ for arbitrary positive values of the parameter $\lambda$. Analogously to the canonical case, the exponents $a_\e$ and $a_m$ determine the static universality class of the corresponding infinite system. Scaling laws which describe the finite-size scaling behaviour of various microcanonical quantities can be derived from the MFSS Assumption.\cite{KaProHue,KastnerDiss}

\section{Consistency of Microcanonical and Canonical Finite-Size Scaling Assumptions}\label{consist}
In this section it is shown that the CFSS assumption (\ref{g_hom}) is a direct consequence of the validity of the MFSS assumption (\ref{s_hom}), and the latter is therefore consistent with the commonly accepted CFSS assumption.

The relation between the microcanonical entropy and the Gibbs free energy can be established from definitions (\ref{sdef}), (\ref{Zdef}), and (\ref{gdef}):
\begin{eqnarray}
\lefteqn{g^{can}(t,h,L^{-1})\,=\,\e_c-T_c(1+t)L^{-d}\ln \bigg\{L^{2d}\,\times}\label{gs_rel_1}\\
&\times&\int\d\es\d m \,\exp\left\{L^d \left[s^{mic}(\es,m,L^{-1})+\frac{hm-\es}{T_c(1+t)}\right]\right\}\bigg\}\;.\nonumber
\end{eqnarray}
As in Eq.\ (\ref{s_split}), the microcanonical entropy is split into a singular and a regular part. To be able to deal with the latter, we make a short excursion to its infinite system analogue. In the thermodynamic limit, $s_{reg}$ {\em per definitionem}\/ does {\em not}\/ determine the critical behaviour of the system. In order to take this into account, an expansion of $s_{reg}$ in powers of $\es$ and $m$ has to be of the form
\begin{equation}
s_{reg}(\es,m)\,=\,\frac{\es}{T_c}+s_-(\es,m)
\end{equation}
where odd powers of $m$ vanish due to the spin inversion symmetry (see Sec.\ \ref{notation}) and the thermodynamically irrelevant constant term of the expansion was set to zero. The linear term $\frac{\es}{T_c}$ fixes the value of the critical temperature
\begin{equation}
\frac{1}{T_c}\,=\,\left.\frac{\partial s(\es,m)}{\partial \es}\right|_{\es,m=0}\;.
\end{equation}
The leading orders in $\es$ and $m$ of the remainder $s_-$ have to be such as not to contribute to the leading asymptotic behaviour of thermodynamic quantities in the vicinity of the critical point $(\es,m)=(0,0)$, which implies
\begin{equation}
\lim_{m\to0}\,\lim_{\es\to0}\,\frac{s_-(\es,m)}{s_{sing}(\es,m)}\,=\,0\;.
\end{equation}
Now we go back to the case of {\em finite}\/ system sizes. Analogously to the infinite system case, for large but finite system sizes $L$ the microcanonical entropy is split into three parts
\begin{equation} \label{sreg_exp}
s^{mic}(\es,m,L^{-1})\,=\,\frac{\es}{T_c}+s^{mic}_{sing}(\es,m,L^{-1})+s^{mic}_-(\es,m,L^{-1})\;.
\end{equation}
Now we {\em define}\/ $s^{mic}_{sing}$ as the part of $s^{mic}$ which, apart from the additive term $\frac{\es}{T_c}$, contains the asymptotically leading terms in $\es$ and $m$ in the vicinity of the critical point $(\es,m)=(0,0)$, i.e., 
\begin{equation} \label{orders}
\lim_{m\to0}\,\lim_{\es\to0}\,\frac{s^{mic}_-(\es,m,L^{-1})}{s^{mic}_{sing}(\es,m,L^{-1})}\,=\,0\;.
\end{equation}
As it is the thus defined $s^{mic}_{sing}$ for which we will obtain a homogeneity relation later on, contact is made to Eq.\ (\ref{s_split}) and the naming ``singular part'' is justified {\em a posteriori}. It is worth noting that $\frac{\es}{T_c}$ is not necessarily the only term in Eq.\ (\ref{sreg_exp}) which is linear in $\es$. The coefficient of the linear term is a function $f$ of the system size $L$ such that it converges towards $\frac{1}{T_c}$ in the limit $L^{-1}\to0$. Thus, for large but finite system sizes we write the term of $s^{mic}$ which is linear in $\es$ as
\begin{equation}
\es f(L^{-1})\,\approx\,\es\left(\frac{1}{T_c}+b L^{-q}\right)
\end{equation}
where $q\in\rz^+$ and $b\in\rz$. The term $\frac{\es}{T_c}$ is treated separately as indicated in (\ref{sreg_exp}), whereas $\es b L^{-q}$, in accordance with Eq.\ (\ref{orders}), is included in the singular part $s^{mic}_{sing}$ of the microcanonical entropy. The exponent $q$, which describes the scaling of the transition temperature with the system size, is determined by the homogeneity relation obtained for $s^{mic}_{sing}$ later on.

We continue our consideration of the canonical Gibbs free energy by inserting (\ref{sreg_exp}) in Eq.\ (\ref{gs_rel_1}) and making use of the geometric series. This yields
\begin{eqnarray}
\lefteqn{g^{can}(t,h,L^{-1})\,=\,\e_c-T_c(1+t)L^{-d}\ln \Bigg\{L^{2d}\,\times}\label{gs_rel_2}\\
&&\times\,\int\d\es\d m \,\exp\left\{L^d \left[s^{mic}_- +s^{mic}_{sing}+\frac{1}{T_c}(h m+t\es)\sum_{k=0}^\infty(-t)^k\right]\right\}\Bigg\}\nonumber
\end{eqnarray}
where, for notational convenience, some functional dependencies have been omitted. The radius of convergence of the geometric series restricts the validity of (\ref{gs_rel_2}) to values of $|t|<1$. Since our intention is to show consistency of finite-size scaling assumptions which hold asymptotically in the vicinity of the critical point, our further proceeding will be to approximate Eq.\ (\ref{gs_rel_2}) for small $t$, $h$, and $L^{-1}$, well within the interval of convergence. To this purpose, higher orders in $t$ and $h$ are dropped to obtain an asymptotic expression for the singular part of the canonical Gibbs free energy
\begin{equation} \label{gs_rel_3}
g^{can}_{sing}(t,h,L^{-1})\simeq-T_cL^{-d}\ln\!\int\!\d\es\d m\,\exp\left\{L^d \left[s^{mic}_- +s^{mic}_{sing}+\frac{h m+t\es}{T_c}\right]\right\}
\end{equation}
for $t,h\approx 0$, where the physically irrelevant additive terms constant in $t$ and $h$ have been omitted. For large enough system sizes $L$, we argue that the maximum of the integrand is located close enough to the critical point $(\es,m)=(0,0)$ such that Laplace's method\cite{Laplace} allows to drop higher orders in $\es$ and $m$ in the integrand of (\ref{gs_rel_3}). Equation (\ref{orders}) implies that this is achieved by neglecting $s^{mic}_-$ in the integrand of (\ref{gs_rel_3}). Thus we obtain
\begin{eqnarray}
\lefteqn{g^{can}_{sing}(t,h,L^{-1})\,\simeq}\label{gs_rel_4}\\
& \simeq & -T_cL^{-d}\ln\int\d\es\d m\,\exp\left\{L^d \left[s^{mic}_{sing}(\es,m,L^{-1})+\frac{1}{T_c}(h m+t\es)\right]\right\}\nonumber
\end{eqnarray}
for $t,h,L^{-1}\approx 0$.

Our next step will be to write down a similar asymptotic equality for $\lambda^{-1}g_{sing}^{can}(\lambda^{a_t}t,\lambda^{a_h}h,\lambda^{1/d}L^{-1})$ in such a way that we can deduce a condition on $s_{sing}^{mic}$ which leads to asymptotic equality of this expression with Eq.\ (\ref{gs_rel_4}) and which therefore implies validity of the CFSS assumption (\ref{g_hom}). From (\ref{gs_rel_4}) we can write
\begin{eqnarray}
\lefteqn{\lambda^{-1}g_{sing}^{can}(\lambda^{a_t}t,\lambda^{a_h}h,\lambda^{1/d}L^{-1})\,\simeq\,-T_cL^{-d}\ln\int\d\es\d m\,\times}\\
& \times & \exp\left\{L^d \left[\lambda^{-1}s^{mic}_{sing}(\es,m,\lambda^{1/d}L^{-1})+\frac{1}{T_c}\left(h\lambda^{a_h-1}m+t\lambda^{a_t-1}\es\right)\right]\right\}\;.\nonumber
\end{eqnarray}
Substitution of the integration variables $\es \longrightarrow \lambda^{1-a_t}\es$ and $m \longrightarrow \lambda^{1-a_h}m$ yields
\begin{eqnarray}
\lefteqn{\lambda^{-1}g_{sing}^{can}(\lambda^{a_t}t,\lambda^{a_h}h,\lambda^{1/d}L^{-1})\,\simeq\,-T_cL^{-d}\ln\int\d\es\d m\,\times}\label{gs_scal}\\
& \times & \exp\left\{L^d \left[\lambda^{-1}s^{mic}_{sing}(\lambda^{1-a_t}\es,\lambda^{1-a_h}m,\lambda^{1/d}L^{-1})+\frac{1}{T_c}(h m+t\es)\right]\right\}\;.\nonumber
\end{eqnarray}
where an additive term $-T_cL^{-d}\ln\lambda^{2-a_t-a_h}$ was dropped, since constants in $t$ and $h$ are not included in the singular part $g_{sing}^{can}$ of the Gibbs free energy. Equating (\ref{gs_rel_4}) and (\ref{gs_scal}) shows that validity of the MFSS assumption (\ref{s_hom}) implies validity of the CFSS assumption (\ref{g_hom}), by which the proof is completed.

\section{Conclusion}\label{concl}
In this paper, we have shown consistency of the MFSS Assumption (\ref{s_hom}) and the CFSS Assumption (\ref{g_hom}), which puts previous papers on MFSS on a firmer footing. As the consistency is shown in an asymptotic sense for large system sizes $L$, the result does not necessarily imply that both Finite-Size Scaling theories work equally well for given finite system-sizes. In fact, it might be the case that in one ensemble or the other, the Finite-Size Scaling region may be reached for smaller system sizes and the quantities under investigation may converge faster towards their infinite system value.

It is worth noting that the MFSS Assumption (\ref{s_hom}) also includes the case where $s^{mic}_{sing}$ is independent of one or more of the variables indicated. This leads to the astonishing observation that even for a microcanonical entropy which is {\em independent of the system size}, canonical quantities show system size dependence according to the CFSS laws (see ref.\ \cite{Trivial} for a detailed discussion). This, however, does not seem to be the relevant case for the investigation of critical phenomena.

\section*{Acknowledgement}
We would like to thank Alfred H\"uller for many valuable discussions.

\begin{appendix}
\section{Reduced Microcanonical or Constant Energy Ensemble}\label{reduc}
In contrast to the above, in the vast majority of papers on the investigation of phase transitions of magnetic systems in the microcanonical ensemble (see references in \cite{KastnerDiss} for a detailed list), the reduced microcanonical partition function
\begin{equation}\label{Ored_def}
\Omega^{red}(\es,L^{-1})\,:=\,\int \d m\,\Omega(\es,m,L^{-1})
\end{equation}
is considered, where the range of integration is $\rz$. From $\Omega^{red}$, the reduced microcanonical entropy
\begin{equation}\label{sred_def}
s^{red}(\es,L^{-1})\,:=\,L^{-d}\ln\Omega^{red}(\es,L^{-1})
\end{equation}
can be derived, which, of course, contains less information than $s^{mic}$. An extrapolation $L\to\infty$ of $s^{red}$ towards the thermodynamic limit allows merely for an estimation of the {\em zero-field} properties of the infinite system under consideration. This becomes obvious from
\begin{eqnarray}
\lefteqn{\lim_{L\to\infty}s^{red}(\es,L^{-1})\,=\,\lim_{L\to\infty}L^{-d}\ln\int \d m\, \exp\left\{L^d s^{mic}(\es,m,L^{-1})\right\}\,=}\nonumber\\
& = & \sup_m s(\es,m)\;=\;\left.\sup_m \left\{s(\es,m)-\frac{h m}{T_c(1+t)}\right\}\right|_{\frac{h}{T_c(1+t)}=0}\,=:\nonumber\\
& =: & \left.\hat{s}(\es,{\textstyle \frac{h}{T_c(1+t)}})\right|_{\frac{h}{T_c(1+t)}=0}
\end{eqnarray}
where
\begin{equation}
s(\es,m)\,:=\,\lim_{L\to\infty}s^{mic}(\es,m,L^{-1})
\end{equation}
is the entropy of the infinite system and $\hat{s}$ is a so-called Massieu function.\cite{Callen}

In order to make contact with papers like ref.~\cite{BruWil}, we want to explicitely transfer our main result to the reduced microcanonical entropy by showing consistency of the homogeneity relation
\begin{equation}\label{sred_hom}
s^{red}(\es,L^{-1})\,\simeq\,\lambda^{-1}s^{red}(\lambda^{a_\e}\es,\lambda^{1/d}L^{-1})
\end{equation}
with the CFSS Assumption (\ref{g_hom}). This can be achieved by showing that (\ref{sred_hom}) is a direct consequence of the MFSS Assumption (\ref{s_hom}).

From (\ref{sdef}), (\ref{sreg_exp}), (\ref{Ored_def}), and (\ref{sred_def}), we obtain
\begin{equation}
s^{red}(\es,L^{-1})\,=\,\es+L^{-d}\ln\int\d m\,\exp\left\{ L^d \left[ s^{mic}_- +s^{mic}_{sing}\right]\right\}\;.
\end{equation}
Due to Eq.\ (\ref{orders}), for large system sizes $L$, Laplace's method\cite{Laplace} allows to drop the term $s^{mic}_-$ in the exponent, and the asymptotic relation
\begin{equation}
s^{red}(\es,L^{-1})\,\simeq\,\es+L^{-d}\ln\int\d m\,\exp\left\{ L^d s^{mic}_{sing}(\es,m,L^{-1})\right\}
\end{equation}
is obtained for $\es,L^{-1}\approx 0$. Making use of the MFSS Assumption (\ref{s_hom}) yields
\begin{equation}
s^{red}(\es,L^{-1})\simeq\es+L^{-d}\ln\lambda^{-a_m}\!\!\int\!\d m\exp\left\{\lambda^{-1}L^d s^{mic}_{sing}(\lambda^{a_\e}\es,m,\lambda^{1/d}L^{-1})\right\}
\end{equation}
where a substitution of the integral variable $m \longrightarrow \lambda^{-a_m}m$ was performed. Neglecting the terms constant in $\es$, a homogeneity relation for the singular part of the reduced microcanonical entropy
\begin{eqnarray}
\lefteqn{s^{red}_{sing}(\es,L^{-1})\,\simeq}\nonumber\\
&\simeq & \lambda^{-1}\left(\lambda^{1/d}L^{-1}\right)^d \ln\int\d m\,\exp\left\{\left(\lambda^{1/d}L^{-1}\right)^{-d}s^{mic}_{sing}(\lambda^{a_\e}\es,m,\lambda^{1/d}L^{-1})\right\}\nonumber\\
& \simeq & \lambda^{-1}s^{red}_{sing}(\lambda^{a_\e}\es,\lambda^{1/d}L^{-1})
\end{eqnarray}
valid for $\es,L^{-1}\approx 0$, is obtained as a direct consequence of the MFSS Assumption (\ref{s_hom}).
\end{appendix}


\end{document}